\documentclass{appolb}
\usepackage{graphicx}
\usepackage{booktabs}
\usepackage{braket}
\usepackage{amsmath}
\usepackage{tikz}
\newcommand{\kaon}{\mathrm{K}^0}
\newcommand{\akaon}{\bar{\mathrm{K}}^0}

\newcommand{\Ks}{\mathrm{K_S}}
\newcommand{\Kl}{\mathrm{K_L}}

\newcommand{\Kp}{\mathrm{K_{+}}}
\newcommand{\Km}{\mathrm{K_{-}}}

\newcommand{\ppm}{\pi^{+}\!\pi^{-}\!}
\begin{document}
\title{Tests of discrete symmetries and quantum coherence with neutral kaons at the KLOE-2 experiment
  \thanks{Presented at the 2$^{\text{nd}}$ Jagiellonian Symposium on Fundamental and Applied Subatomic Physics}%
}
\author{Aleksander Gajos\\
  on behalf of the KLOE-2 collaboration  
  \address{%
    Institute of Physics, Jagiellonian University,\\
    ul.\ \L{}ojasiewicza 11, 30-348 Cracow, Poland\\
    aleksander.gajos@doctoral.uj.edu.pl
    }
  \\
  }
\maketitle
\begin{abstract}
The KLOE-2 detector records decays of quantum-entangled pairs of neutral kaons produced in decays of $\phi$ mesons provided by the DA$\Phi$NE accelerator at the Laboratori Nazionali di Frascati, Italy. This system allows for a broad range of studies of fundamental discrete symmetries including tests which are only feasible with entangled neutral mesons. This work reports on the Lorentz non-invariance and CPT violation searches with the $\phi\to\Ks\Kl\to\pi^+\pi^-\pi^+\pi^-$ process in the framework of the Standard Model Extension. Moreover, status and results of quantum coherence tests with the same process are discussed. Finally, the status of ongoing direct tests of T and CPT in neutral kaon transitions is presented. For each of the reported studies, perspectives are discussed for the KLOE-2 experiment, which is presently taking data.
\end{abstract}
  
\section{Introduction}
Systems of flavoured neutral mesons have long proven to constitute an excellent ground for studies of the fundamental discrete symmetries of Nature as conservation of the latter can be easily translated to constraints on the properties of such systems. Among them, neutral kaons are of special importance as the value of the mass difference between $\kaon$ and $\akaon$ state enables the use of quantum-entangled pairs of neutral K mesons to observe a range of quantum intereference phenomena. Such interferometric measurements may also be used to design tests of discrete symmetries and quantum mechanics  as has been shown by the KLOE experiment, which to date is the sole experimental setup where $\kaon\akaon$ pairs are available in a quantum entangled state, provided by the DA$\Phi$NE $\phi$-factory. Presently, the physics programme is continued by the KLOE-2 detector. This work reports on the recent tests of both discrete symmetries and quantum coherence performed by KLOE and the perspectives for KLOE-2 results.

\section{The KLOE-2 setup at the DA$\Phi$NE $\phi$-factory}
The KLOE-2 detector records decays of $\phi$ mesons produced in $e^+e^-$ collisions by the DA$\Phi$NE collider operating in the National Laboratories of Frascati in Italy. The detection setup common for KLOE and KLOE-2 consists of a cylindrical drift chamber (DC)~\cite{Adinolfi:2002uk} and a sampling electromagnetic calorimeter (EMC) surrounding it with a coverage of 98\% of 4$\pi$~\cite{Adinolfi:2002zx}. The DC uses a light gas mixture to minimize kaon regeneration and its size with an outer radius of 2\:m is driven by the need to detect a large part (40\%) of decays of the $\Kl$ mesons, whose mean path in KLOE is about 3.4\:m. The calorimeter provides good timing resolution which is used to reconstruct the $\Kl\to 3\pi^0\to 6 \gamma$ decays based on the $\gamma$ interaction points in the EMC~\cite{Gajos:2015qsa}. Additionally, the KLOE-2 uses a novel Inner Tracker device based on 4 layers of Cylindrical Gas Electron Multiplier detector aiming at improving tracking and vertexing resolution close to the interaction point~\cite{kloe_inner_tracker}.

\section{Search for CPT and Lorentz symmetry violation with the $\phi\to\Ks\Kl\to\pi^+\pi^-\pi^+\pi^-$ process}
In the framework of the Standard Model Extension (SME)~\cite{Kostelecky:2001ff} and the anti-CPT theorem~\cite{Greenberg:2002uu}, violation of CPT must be associated to Lorentz non-invariance which allows to test CPT by searching for Lorentz symmetry-violating effects. In the neutral K meson system, the latter would be manifested by a dependence of the usual CPT-related $\delta$ parameter~\cite{interf_handbook} on the Lorentz factor $\gamma_K$ and momentum direction $\vec{\beta}_K$ of a kaon:
\begin{equation}
  {\delta_K} \simeq i\: sin \phi_{SW} e^{i\phi_{SW}}{\gamma_K}({\Delta a_0} - {\vec{\beta_K}}{\Delta \vec{a}})/ \Delta m,
  \label{eq:delta-vs-p}
\end{equation}
where $\Delta a_{\mu}$ denote the parameters of the SME Lagrangian part and $\phi_{SW}$ is the so-called superweak phase.

Experimentally, $\delta_K$ as a function of the kaon momentum direction can be extracted from a double decay amplitude for quantum-entangled $\Ks\Kl$ pairs with both kaons decaying into the $\ppm$ final state:
 \begin{equation}
   I(\Delta\tau) \sim e^{-\Gamma |\Delta\tau|} \Big[ |\eta_{1}|^2 e^{\frac{\Delta\Gamma}{2} \Delta\tau} + |\eta_{2}|^2 e^{-\frac{\Delta \Gamma}{2}\Delta\tau} -2 \Re e \Big( \eta_{1} \eta_{2}^* e^{-i\Delta m \Delta\tau} \Big) \Big],
   \label{eq:idt}
 \end{equation}
where the $\eta_{1(2)}$ amplitudes are dependent on $\delta(\vec{p_K})$ and $\Delta \tau$ denotes time difference between both kaon decays.

To this end, $\phi\to\Ks\Kl\to\pi^+\pi^-\pi^+\pi^-$ events were selected from the 1.7 fb$^{-1}$ of data recorded by KLOE. To account for direction dependence, the events were divided into two angular subsamples based on the projection of the momentum of the more energetic of two kaons on the $\phi$ direction in the laboratory reference frame. Consequently, daily movement of the laboratory was expressed in the frame of fixed stars and the data sample was further split into 4 intervals of sidereal time. For each of the eight obtained subsamples of $\Ks\Kl\to\pi^+\pi^-\pi^+\pi^-$ events, the double decay rate~(Eq.~\ref{eq:idt}) was fit simultaneously in order to extract the $\Delta a_{\mu}$ SME parameters of the kaon sector. The results, presented in Table~\ref{tab:lorentz}, reach the expected sensitivity at the level of $10^{-18}$ GeV~\cite{Babusci:2013gda}, which is several orders of magnitude more precise than results obtained with other neutral meson systems~\cite{PhysRevLett.116.241601}. Moreover, further improvement is expected with the KLOE-2 detector due to its goal integrated luminosity of 5 fb$^{-1}$ as well as the enhanced tracking capabilities. Expected KLOE-2 statistical uncertainty is shown in Table~\ref{tab:lorentz}.

\begin{table}[h]
  \caption{\label{tab:lorentz}Values of the Lorentz symmetry-violating SME parameters measured by the KLOE experiment with 1.7 fb$^{-1}$ of data~\cite{Babusci:2013gda}.
    The rightmost column presents statistical uncertainty expected with data taken by the KLOE-2 experiment and 5~fb$^{-1}$ of data.
  }
  \begin{center}
    \begin{tabular}{ccc}
      \toprule
      \begin{minipage}{4em}\centering SME\\Parameter\end{minipage} & KLOE measurement & \begin{minipage}{9em}\centering Expected KLOE-2\\uncertainty\end{minipage}\\
      \midrule
      $\Delta a_0$ & $(-6.0\pm 7.7_{stat}\pm 3.1_{syst})\times 10^{-18}\:\mathrm{GeV}$  &  $\pm 2.2_{stat}\times 10^{-18}\:\mathrm{GeV}$ \\
      $\Delta a_X$ & $\phantom{-}(0.9\pm 1.5_{stat}\pm 0.6_{syst})\times 10^{-18}\:\mathrm{GeV}$ & $\pm 0.4_{stat}\times 10^{-18}\:\mathrm{GeV}$ \\
      $\Delta a_Y$ & $(-2.0\pm 1.5_{stat}\pm 0.5_{syst})\times 10^{-18}\:\mathrm{GeV}$ & $\pm 0.4_{stat}\times 10^{-18}\:\mathrm{GeV}$ \\
      $\Delta a_Z$ & $\phantom{-}(3.1\pm 1.7_{stat}\pm 0.6_{syst})\times 10^{-18}\:\mathrm{GeV}$ & $\pm0.5_{stat}\times 10^{-18}\:\mathrm{GeV}$ \\
      \bottomrule
    \end{tabular}
  \end{center}
\end{table}

\section{Search for quantum decoherence with $\phi\to\Ks\Kl\to\pi^+\pi^-\pi^+\pi^-$}
Interferometric studies of the double decay rate for a pair of neutral kaons decaying into $\ppm\ppm$ also enable a test of Quantum Mechanics through a measurement of decoherence level for the entangled K mesons. Such decoherence, which could occur after the $\phi\to\kaon\akaon$ decay according to the hypothesis of Furry~\cite{furry}, can be parametrised by inserting a $\zeta$ decoherence parameter before the interference term in the double decay amplitude~(for the $\kaon\akaon$ basis):
\begin{equation}
  \label{eq:i_zeta}
  \begin{split}
    I(\ppm, & \ppm,\Delta t) =
    \frac{N}{2}\Big [ |\langle \ppm,\ppm | \kaon\akaon(\Delta t)\rangle|^2 + |\langle \ppm,\ppm | \akaon\kaon(\Delta t)\rangle|^2  \\
    &- (1-\zeta_{0\bar{0}}) \cdot 2 \Re \left( \langle \ppm,\ppm | \kaon\akaon(\Delta t)\rangle \langle \ppm,\ppm | \akaon\kaon (\Delta t)\rangle ^*\right) \Big],
  \end{split}
\end{equation}
where $N$ is a normalization constant. Similarly, a $\zeta_{SL}$ parameter is defined for the $\left\{\Ks,\Kl\right\}$ basis.

Any possible quantum decoherence would therefore manifest itself as a reduction of the dip in the double decay rate for  $\Delta t \approx 0$. This is caused by destructive interference which prevents both kaons from decaying into the same final state at the same moment.

\begin{figure}[htb]
\centering
  \includegraphics[height=5cm]{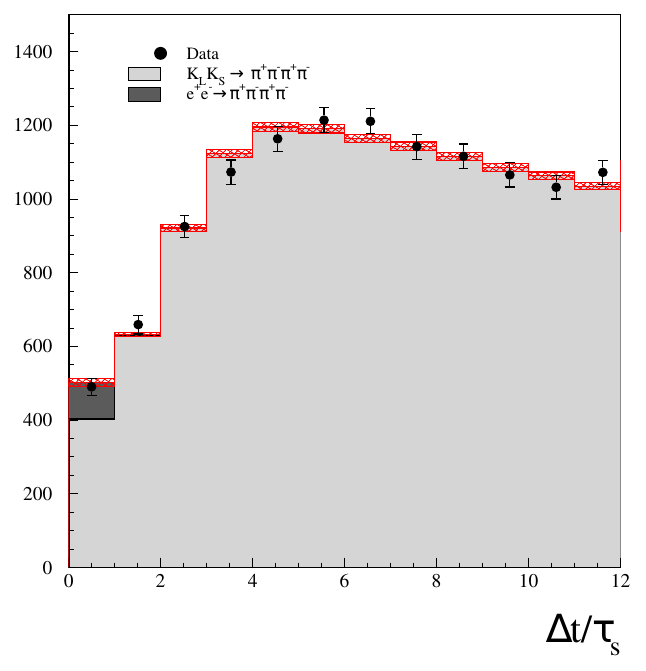}\label{fig:decoh_old}
  \hfill
  \includegraphics[height=5cm]{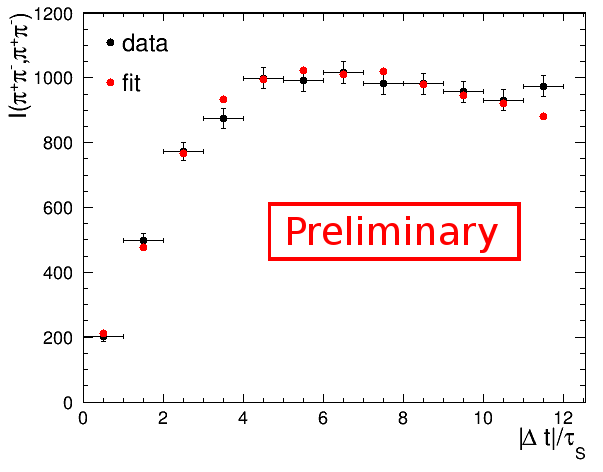}\label{fig:decoh_new}
\caption{\label{fig:decoh}Fits to double decay rates for $\Ks\Kl\to\pi^+\pi^-\pi^+\pi^-$ as function of time difference between kaon decays obtained with published KLOE data analysis (left)~\cite{decoherence} and a preliminary result of refined analysis (right) where an improved reproduction of the interference pattern for $\Delta t \approx 0$ is visible. }
\end{figure}

Both $\zeta_{SL}$ and $\zeta_{0\bar{0}}$ (for which sensitivity is naturally higher due to CP suppression in the decay channel) were extracted by KLOE with a fit to the double decay rate for a sample of $\phi\to\Ks\Kl\to\pi^+\pi^-\pi^+\pi^-$ decays displayed in the left panel of Fig.~\ref{fig:decoh}. The measurement yielded no observation of decoherence as the following values were obtained~\cite{decoherence}:
\begin{eqnarray}
  \zeta_{0\bar{0}} & = & (1.4 \pm 9.5_{stat} \pm 3.8_{syst}) \times 10^{-7},\\
  \zeta_{SL} & = & (0.3 \pm 1.8_{stat} \pm 0.6_{syst}) \times 10^{-2}.
\end{eqnarray}
Even though the above results significantly improve previous measurements with neutral K~\cite{PhysRevD.60.114032} and B mesons~\cite{PhysRevLett.99.131802},
the factors limiting sensitivity comprise statistical uncertainty as well as $\Delta t$ resolution of the interference region ($\Delta t \approx 0$), and a residual background component of non-resonant $e^+e^-\to\ppm\ppm$. The same dataset is therefore presently reanalised with a refined event selection which allows for a better reproduction of the interference pattern as shown in the preliminary distribution in the right panel of Fig.~\ref{fig:decoh}. Additionally, KLOE-2 data will bring a further improvement due to the use of new Inner Tracker subdetector.

\section{Test of time-reversal and CPT symmetry in neutral kaon transitions with $\phi\to\Ks\Kl\to \pi e \nu \: 3\pi^0\:(2\pi)$ decays}
A novel concept of direct tests of discrete symmetries in neutral kaon transitions is presently pursued at KLOE-2 for the T and CPT symmetries. The test uses kaon transitions between their pure flavour $\{\kaon,\akaon\}$ and CP-definite states $\{\Kp,\Km\}$ and is based on a comparison between rates of processes like $\kaon\to\Km$ and their time-reversal conjugates obtained by an exchange of initial and final states~\cite{Bernabeu-t,Bernabeu-cpt}.

\begin{figure}[htb]
  \centering
  \includegraphics[width=0.48\textwidth]{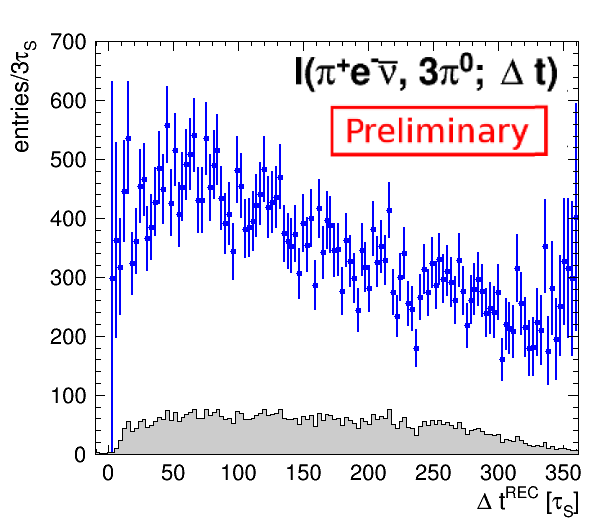}\label{fig:dt1}
  \includegraphics[width=0.48\textwidth]{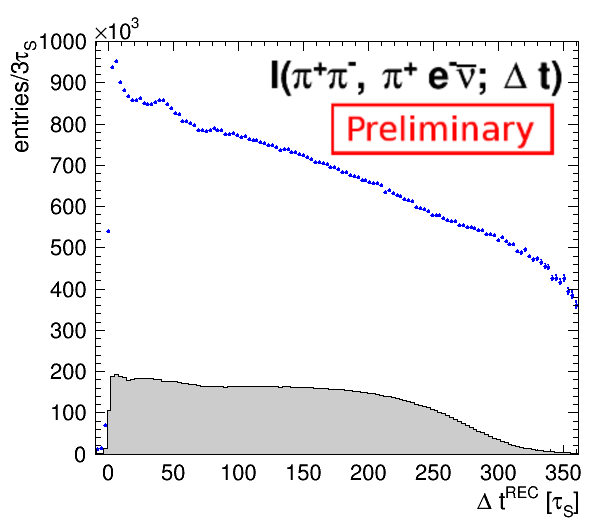}\label{fig:dt2}
\caption{Distributions of double decay rates as a function of kaon decay times difference for two chosen processes entering the determination of the asymmetries used in the direct T and CPT tests. Filled grey histograms present raw spectra while the ones corrected for selection efficiency are marked with blue points.}
\label{fig:dt}
\end{figure}

The kaon states are identified by semileptonic decays $\kaon\to\pi^-e^+{\nu}$ and $\akaon\to\pi^+ e^- \bar{\nu}$ as well as hadronic decays into two ($\Kp\to\ppm$) and three pions ($\Km\to 3\pi^0$). Initial state of the kaon in the compared transitions is inferred from the decay of its quantum-entangled partner, a technique which is only possible at DA$\Phi$NE in the case of neutral K mesons.

The $\Kl\to 3\pi^0\to 6 \gamma$ decay involving only neutral particles is reconstructed using solely information on the photon interactions in the KLOE EMC with a trilateration-based technique specially devised for this study~\cite{Gajos:2015qsa}. The event selection and reconstruction procedures, which will be applied to the dataset of KLOE-2, are currently being prepared and tested with the data already recorded by KLOE.\ Figure~\ref{fig:dt} presents preliminary distributions of rates of events corresponding respectively to the transitions $\kaon\to\Km$ and $\Km\to\akaon$, whose ratio $R^{exp}_{2,CPT}(\Delta t) \sim P[\kaon\to\Km]/P[\Km\to\akaon]$ constitutes one of the observables of the CPT test\footnote{The reader is referred to~\cite{Bernabeu-t} and~\cite{Bernabeu-cpt} for a complete definition of observables for the T and CPT tests respectively.}.

\begin{figure}[htb]
  \centering
  \includegraphics[width=0.8\textwidth]{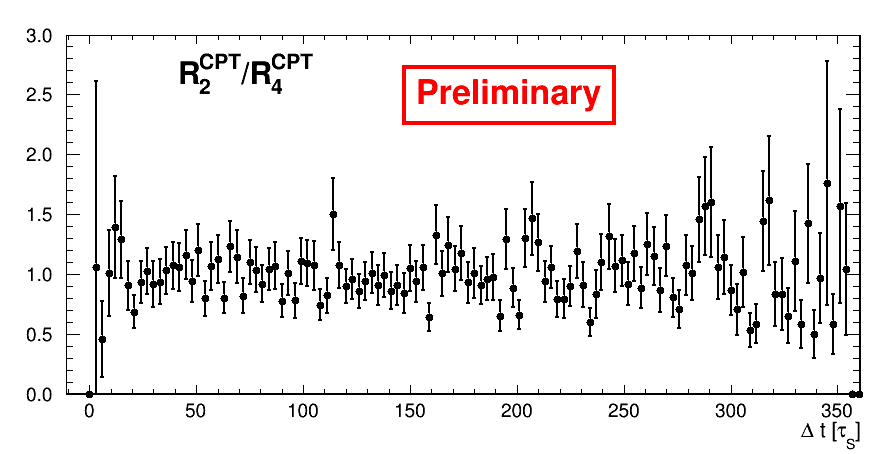}
  \caption{Preliminary distribution of the CPT-asymmetric double ratio $R^{exp}_{2,CPT}/R^{exp}_{4,CPT}$ obtained with the KLOE data.}
  \label{fig:r2r4}
\end{figure}

An especially robust observable of the CPT symmetry test in transitions of neutral kaons is a ratio of the aforementioned CPT-violating ratios $R^{exp}_{2,CPT}$ and $R^{exp}_{4,CPT}(\Delta t) \sim P[\akaon\to\Km]/P[\Km\to\kaon]$ expressed as a function of the kaons' decay time difference. It was shown that the asymptotic value of this ratio in the region of $\Delta t \gg \tau_{S}$ (where $\tau_S$ denotes lifetime of $\Ks$) is related to the CPT-violating parameters in the following way~\cite{Bernabeu-cpt}:
\begin{equation}
  \label{eq:r2r4} 
  \frac{R_{2,CPT}^{exp}(\Delta t \gg \tau_S)}{R_{4,CPT}^{exp}(\Delta t \gg \tau_S)} = 1 - 8\Re\delta -8\Re x_{-} .
\end{equation}

A measurement of the asymptotic value of this CPT-violation sensitive double ratio is one of the objectives of KLOE-2. Figure~\ref{fig:r2r4} presents a preliminary result obtained with the KLOE dataset with an integrated luminosity of 1.7 fb$^{-1}$.

\section{Summary and perspectives}
Although the KLOE detector has already provided numerous contributions to the tests of fundamental discrete symmetries such as the CPT and Lorentz symmetries and the foundations of Quantum Mechanics through decoherence searches, KLOE data are still exploited to provide more precise results as in the case of the decoherence parameter estimation. Moreover, a novel test of time reversal and CPT symmetries is in progress using the data collected by the KLOE-2 detector, which is presently in operation. KLOE-2 will bring not only larger statistics of at least 5 fb$^{-1}$ but also improved event reconstruction due to new detector components such as the Inner Tracker. These factors are expected to improve sensitivity of all the reported tests.

\section*{Acknowledgements}
We warmly thank our former KLOE colleagues for the access to the data collected during the KLOE data taking campaign.
We thank the DA$\Phi$NE team for their efforts in maintaining low background running conditions and their collaboration during all data taking. We want to thank our technical staff:
G.F.~Fortugno and F.~Sborzacchi for their dedication in ensuring efficient operation of the KLOE computing facilities;
M.~Anelli for his continuous attention to the gas system and detector safety;
A.~Balla, M.~Gatta, G.~Corradi and G.~Papalino for electronics maintenance;
C.~Piscitelli for his help during major maintenance periods.
This work was supported in part
by the Polish National Science Centre through the Grants No.\
2013/08/M/ST2/00323,
2013/11/B/ST2/04245,
2014/14/E/ST2/00262,\\
2014/12/S/ST2/00459,
2016/21/N/ST2/01727.

\section*{References}


\end{document}